**Bridge Deck Delamination Segmentation Based on Aerial Thermography Through Regularized Grayscale Morphological Reconstruction and Gradient Statistics**

**Chongsheng Cheng**
The Durham School of Architectural Engineering and Construction
University of Nebraska-Lincoln
113 Nebraska Hall, Lincoln, NE 68588-0500
E-mail: cheng. chongsheng@huskers.unl.edu

**Zhexiong Shang**
The Durham School of Architectural Engineering and Construction
University of Nebraska-Lincoln
113 Nebraska Hall, Lincoln, NE 68588-0500
E-mail: szx0112@huskers.unl.edu

**Zhigang Shen, Corresponding Author**
The Durham School of Architectural Engineering and Construction
University of Nebraska-Lincoln
113 Nebraska Hall, Lincoln, NE 68588-0500
E-mail: shen@unl.edu

**ABSTRACT**
Environmental and surface texture-induced temperature variation across the bridge deck is a major source of errors in delamination detection through thermography. This type of external noise poises a significant challenge for conventional quantitative methods such as global thresholding and k-means clustering. An iterative top-down approach is proposed for delamination segmentation based on grayscale morphological reconstruction. A weight-decay function was used to regularize the reconstruction for regional maxima extraction. The mean and coefficient of variation of temperature gradient estimated from delamination boundaries were used for discrimination. The proposed approach was tested on a lab experiment and an in-service bridge deck. The result showed the ability of the framework to handle the non-uniform background situation which often occurred in practice and thus eliminates the need of inferencing the background required by existing methods. The gradient statistics of the delamination boundary in thermal image were indicated as the valid criterion refine the segmentation under proposed framework. Thus, improved performance was achieved compared to the conventional methods. The parameter selection and the limitation of this approach were also discussed.

*Keywords*: Thermography, Delamination segmentation, Bridge deck, Non-uniform thermal background, Grayscale morphological reconstruction, Global threshold, K-means clustering

## 1. INTRODUCTION

NDE method for detecting shallow delamination (less than 10.2 centimeter or 4.0 inches) of bridge deck has been reported in several studies (Dabous et al. 2017; Oh et al. 2012; Omar and Nehdi 2017; Vaghefi et al. 2013). The principle of detection is based on the temperature contrast between the delaminated and sound areas (Hiasa et al. 2017) during the daily solar heating cycles. Although the principle is straight-forward, the quality of thermal images are often affected by unfavorable factors such as inappropriate time window of the image-shots, insufficiently developed temperature contrast, and surface inhomogeneities (Cheng and Shen 2018; Washer et al. 2009; Watase et al. 2015). Therefore, accurately segmenting delamination out of a raw thermal image remains a challenge. To address the challenge, several quantitative methods have been developed to process the thermal image based on temperature contrast, temperature gradient, and temperature density distribution (Dabous et al. 2017; Ellenberg et al. 2016; Kee et al. 2011; Oh et al. 2012; Omar and Nehdi 2017; Vaghefi et al. 2013). Despite some progresses, these methods suffered degraded performance under spatial temperature variations, which were often referred to as the non-uniform thermal background. This issue was observed in both experimental setup when using a non-uniform excitation heating source (Cheng et al. 2018; Milovanović et al. 2017), and under the natural outdoor environment (Sultan and Washer 2017). The underlying assumption of these methods required a relative uniformed background to represent the "sound" concrete area so that the highest temperature areas were associated with the delamination. However, this assumption cannot always be satisfied (Sultan and Washer 2017). Thus, the delaminated areas represented by regional maxima were often missed. To address the issue, new delamination segmentation approach needs to be developed so it can segment regional maxima under the non-uniform background.

Conventional methods to segment the delamination were mainly based on a bottom-up framework which required the determination of the reference temperature beforehand. Hard thresholding on temperature value or the percentile of temperature variation was adopted by Dabous et al. (2017), Kee et al. (2011), Oh et al. (2012), and Vaghefi et al. (2013) to convert the raw thermal image into a binary image to indicate the delamination by assuming the low temperature or the bottom percentile as the sound area. Due to the observed large temperature variation across the entire bridge deck (Dabous et al. 2017; Oh et al. 2012; Sultan and Washer 2017), the scanned area needed to be split into sub-images to apply the discrimination criterion, and thus no single threshold could be used as the global criterion on each sub-image. Therefore, the threshold method heavily relied on the operator's experience and was time-consuming. The clustering based method was introduced by Omar and Nehdi (2017) that a k-means based clustering model was developed to indicate the delamination area by inspecting on the condition of bridge decks ( i.e. bridge age, temperature contrast, and the dimension of founded surface spalling). Based on the condition of the bridge deck, the number of K could be determined so that the thermal image could be classified into k groups. Since the k-means clustering algorithm is based on the distance metric of density, the sound area was selected as the lowest-mean group during daytime and the highest-mean group during nighttime without considering the spatial information. The region growth method was utilized by Abdel-Qader et al. (2008) from which an automatic process was developed to segment the defected area in the thermal image. The method assumed that the delaminated area shared the highest temperature over the scene and thus used the neighbor temperature deviation difference inside a 9-by-9 pixel window as the criterion for screening the image. Ellenberg et al. (2016) extended the method through using temperature gradient as the threshold criterion. Although the region growth-based segmentation method showed the insensitivity to the non-uniform background, the limitation remained on the seed allocation and growing-and-stop mechanism that was still yielded by the delaminated area to be the global maxima in temperature (Abdel-Qader et al. 2008) or temperature gradience (Ellenberg et al. 2016). In general, these methods relied on determining the reference while the true delamination is more associated with regional maxima and spatial characteristics.

The objective of this study is to develop a top-down segmentation method, using grayscale morphology, to segment the delamination of the bridge deck from the aerial surveyed thermal image under natural environment. An iterative procedure with initial condition is designed for regional maxima extraction based on the grayscale morphological reconstruction and a weight function is selected as a regularizer to constrain the reconstruction procedure for better preservation of the regional maxima. Then the statistical measure from the gradient map is used to discriminate the delamination out of regional maxima. In this paper, the proposed method was first tested on experimental cases using simulated delamination in a concrete slab and then a case study using an actual concrete bridge deck was carried out. The results are compared with existing threshold-based and clustering-based methods.

## 2. BACKGROUND
### 2.1 Temperature Contrast as The Minimum Criterion

The temperature contrast as the criterion for detecting delamination has been widely studied and the lower bond was explicitly discussed in the past experimental studies which defined the minimum criterion. According to ASTM (D4788-03), at least 0.5 ºC difference between the delaminated area and adjacent sound area was recommended. This recommendation only describes the minimum temperature contrast under the ideal condition, but the contrast varies in practice in terms of different delamination size and depth (Washer et al. 2009) as well as the different time window of a day (Kee et al. 2011; Watase et al. 2015). Recently, a comprehensive study was conducted by Hiasa et al. (2017) to evaluate the detectability of delamination detection in terms of different sizes, shapes, and depth using thermography throughout the finite element simulation. The study concluded that with the increase of the delamination area, the temperature contrast increased until it became large enough to be stable (e.g. over 40x40 cm). Also, it showed the temperature contrast reached a positive maximum (the delaminated area was higher than sound area) during the noon to afternoon. Thus, 0.4 ºC was recommended as the lower bond of the contrast for the certain detectability. Sultan and Washer (2017) conducted a reliability analysis of delamination detection by thermography based on the receiver operating characteristics (ROC) of the temperature contrast. It evaluated the effect of spatial temperature variation on the contrast-based thresholding method and 0.6 ºC was suggested as the optimum contrast for in-service bridge decks and 0.8 ºC for a simulated slab with considerable accuracy (80%). Even though an effort has been made for the selection of the rational temperature contrast, as far as the bottom-up framework is concerned, the remaining challenge is to determine the reference without prior awareness of delamination allocations before applying the criterion.

### 2.2 Grayscale Morphologic Reconstruction for Regional Maxima Extraction
Grayscale morphology was derived from mathematic morphology as a useful approach for image processing based on the geometric shape of the object presented in the image (Haralick et al. 1987). It is based on the set theory and consists of the combination of dilation, erosion, and set operations

(e.g. union, intersection, and complement). It has been used for image filtering, edge detection, denoising, region filling, and segmentation (Coster and Chermant 2001; Maragos 1987; Vincent 1993). The applications in civil engineering have been found for concrete material analysis (Coster and Chermant 2001), region recognition from LiDAR data (Cheng et al. 2013; Vosselman 2000), and pavement crack detection (Sun and Qiu 2007; Wei et al. 2009). The advantage of using the morphological operation was to transform the image based on the object's shape so that geometric information could be controlled throughout the manipulation (Rishikeshan and Ramesh 2018).

The grayscale morphological reconstruction is an iterative operation based on dilation and erosion that is often used for the dome (regional maximal) and basin (regional minima) extraction. Given two functions to represent a mask image ($F$) and a marker image ($G$), $G<F$ means $G$ "under" $F$. The reconstruction by dilation is then defined below:

$$R_F^D(G) = D_F^k(G), \text{ when } D_F^k(G) = D_F^{k+1}(G) \quad (1)$$

Which $D_F^1 = (G \oplus B) \wedge F, D_F^k = D_F^1(D_F^{k-1}(G)) \quad (2)$

Where $\wedge$ is the inclusion operation from Set theory and a "pointwise minimization" to ensure that $(G \oplus B)$ is "under" the mask $F$. Thus, the reconstruction by dilation $R_F^D(G)$ can be interpreted as the marker image $G$ that dilates with the structure element $B$ under the mask image $F$ until stability is reached $(D_F^k(G) = D_F^{k+1}(G))$. When the marker image ($G$) equals F-h (a special case that marker image $G$ is the shifted $F$ by offsetting $h$) the reconstruction can be used for dome extraction. Figure 1 illustrated the 1D representation of this procedure. Based on the definition of reconstruction, the key to extract the regional maxima is depended on the selection of $h$ value which defines the contrast between maxima and its surroundings. This property becomes useful for our case since the temperature contrast is the principle to distinguish delamination from the sound area.

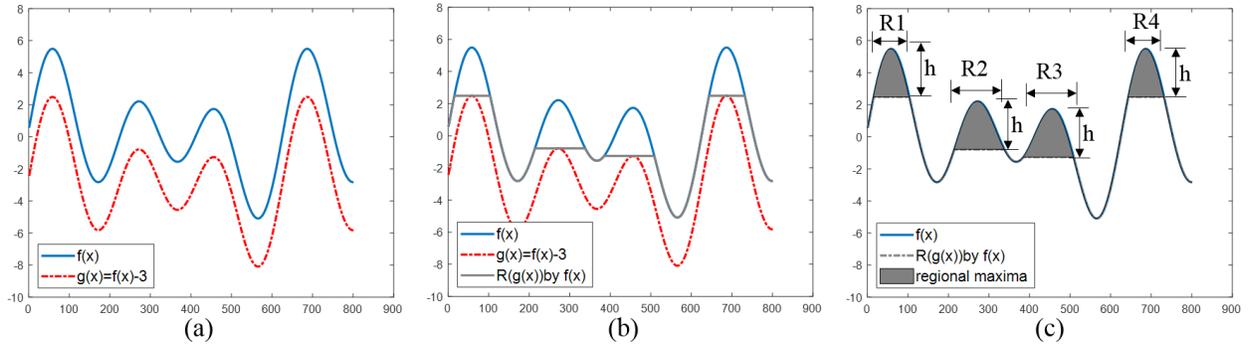

**FIGURE 1. Illustration of grayscale morphologic reconstruction for 1D representation: (a) synthetic 1 D signal f(x) = sin(x)+2*cos(2*x+5) +3*sin(3*x) and g(x) = f(x)-3; (b) reconstruction of g(x) by f(x); (c) regional maxima by difference between f(x) and R(g(x))**

## 3. METHODOLOGY
### 3.1 Delamination Thermal Characteristic, Regional Maxima, and Regularized Morphological Reconstruction

The thermal characteristic of delamination embedded in bridge deck has an implicit definition in the literature which often refers to "hot" or "cold" regions based on the surroundings. Here we link this observed phenomenon to a more rigorous definition of regional extrema in mathematic morphology and from now on we focus on the regional maxima as the "hot" region due to its typicality in practice. The regional maxima defines the region which all values inside is constant and larger than any values on the outward boundary neighbors. Based on the definition, the constant could be calculated by grayscale morphological reconstruction by the certain offset *h*, and the regional maxima are determined by the associated connectivity to its neighbors. The region could be identified as one of the maxima only and if only all existing neighbors are smaller than the region. Thus, with the changing of *h*, the regional maxima could "expand", "merge", and "vanish". With gradual increase of *h*, the regional maxima expand (Fig. 2 top: region 1 and 4 through a to c), merge (Fig. 2b top: region 2 and 3) and vanish (Fig. 2c top).

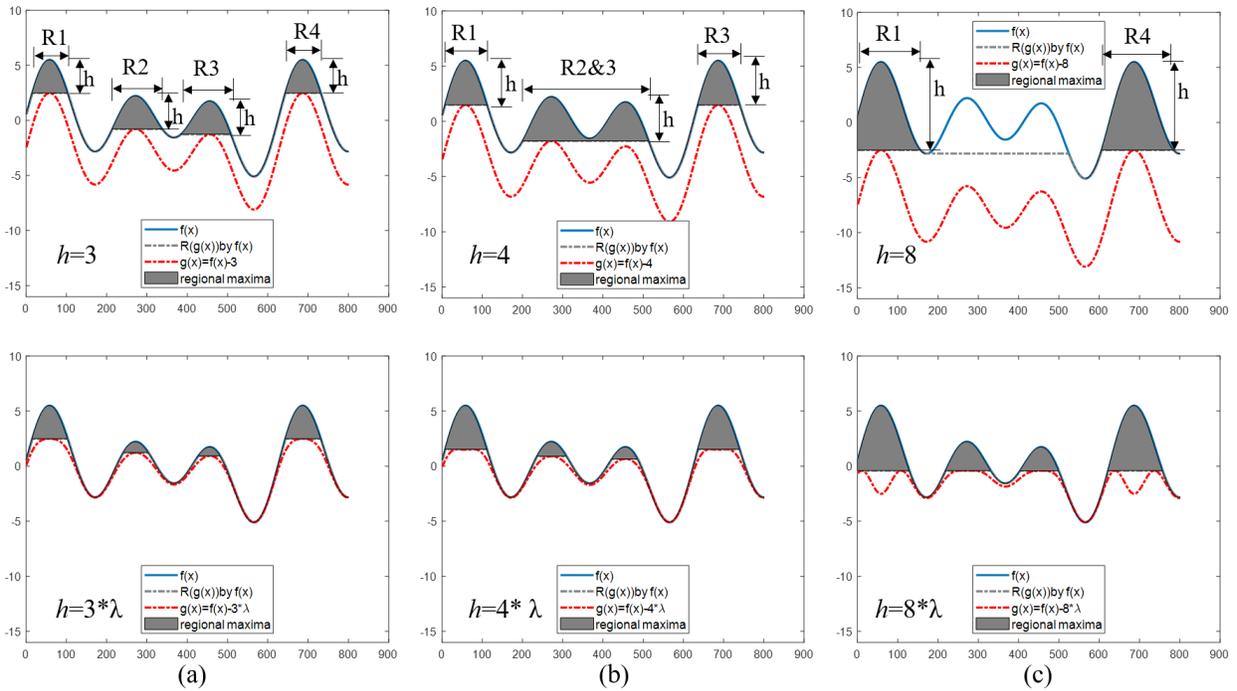

**FIGURE 2. Illustration of "expand", "merge", and "vanish" of regional maxima (top) with different *h* values: four regional maxima with *h* = 3 (a); three regional maxima with *h* = 4 (b); and two regional maxima with *h* = 8 (c). Regularized morphological reconstruction for regional maxima (bottom): Four regional maxima are preserved in the bottom through (a) to (c).**

The general grayscale morphological reconstruction may merge and vanish the regional maxima quickly due to the global offsetting *h* value which may not be a feasible behavior due to the potential large background variation by aerial surveyed thermal images. Thus, we used a cubic weight function to constrain this behavior so that during reconstruction, with a higher hill having a deeper offset; and a lower hill having a shallow offset. The function is then defined as (eq.3):

$$\lambda = w^3 \quad (3)$$

Where *w* is the normalized weight of the original image to the range of [0,1]. The function λ not only preserves the boundary condition that $\lambda(0) = 0$ and $\lambda(1) = 1$, but also performs a decay of weight that a convex function between the 0 and 1. After applying the function λ to the offest *h*, the reconstruction is more regulated so that four regional maxima are preserved in Fig. 2 (bottom).

## 3.2 The Framework

The framework consists of two parts: the generation of regional maxima and the discrimination of delamination out of regional maxima (Fig. 3). The hypothesis is that the delamination is a subset of regional maxima so that the identified regional maxima needs to be screened by the feature of delamination to ensure the bijective property. The raw image is first smoothed by anisotropic diffusion procedure (Weickert 1998) to have $T_s$ and gradient map $G_s$. The first reconstruction $I_0$ is the general grayscale morphological reconstruction by the global offset $h_{in}$ which is to satisfy the minimum criterion of temperature contrast (Section 2.1). Then the regulerized reconstruction is conducted by a small stepsize Δ to let the regional maxima expand. The iteration will stop under two criterions: 1) the maximum steps are reached which is defined by $n = \max(temperature\ contrast)/\Delta$; 2) the criterion satisfy the maximally stable extremal regions by Matas et al. (2004). On each iteration, the regional maxima ($R_n$) have been detected based on the definition (Section 3.1).

The discrimination part utilizes the estimated gradient information of delamination from the gradient map $G_s$ to screen the generated regional maxima. Firstly, the delamination gradient is estimated from gradient map $G_s$ by the clustering method (e.g. k-means) and then the statistical measures $D_g$ (mean and coefficient of variation) are extracted as feature templates. The assumption is based on the boundaries of extracted regional maxima on each iteration to be classified as delamination, only if they match the information to the boundaries of true delamination ($D_g$). The gradient mean represents the mean value of gradient strength which is selected based on the observation that the gradient strength is consistent for delamination at a certain depth and time of day. The coefficient of variation (defined as $V_{var} = \delta_{std}/M_{grad}$) measures the extent of variability in relation to the mean of a distribution which we assume this variability is similar within the true boundaries of delamination and distinct from non-delamination regions. The discrimination is made between sub regions $r_i$ from $R_n$ and $D_g$. $R'_n$ includes the screened regional maxima that are identical to delamination. The final segmentation of delamination is the union of each $R'_n$.

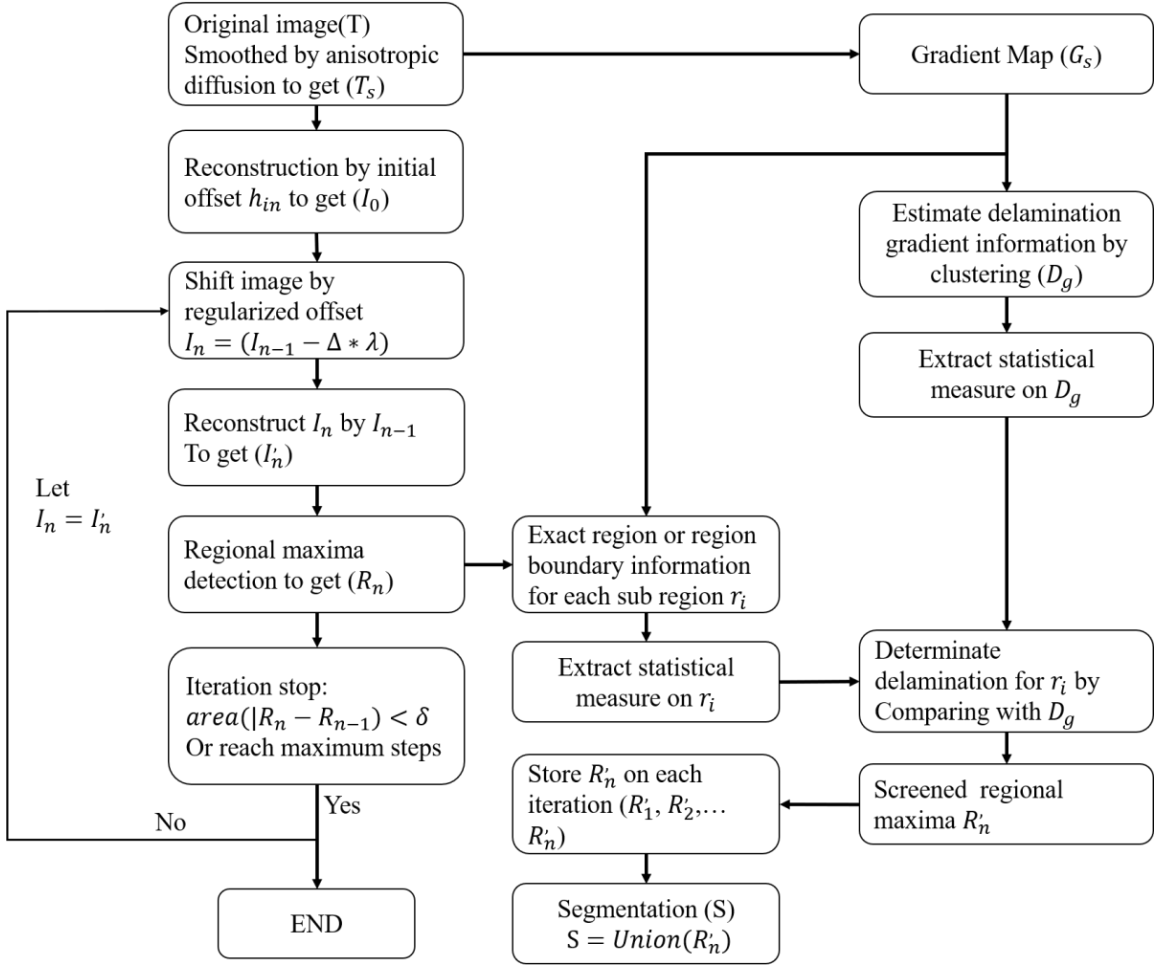

**FIGURE 3. The data flow**

## 4. EXPERIMENTAL TEST

An experimental study was conducted using mimicked delamination in a reinforced concrete slab outdoors, in sunny weather. The reinforced concrete slab is illustrated in Fig. 4. A thermal camera was used to collect the surface temperature data on August 22, 2018. The raw image (Fig. 4d) was taken at 3 pm. Table 1 shows the camera configurations. During the data processing, the initial offset $h_{in} = 0.5°C$ and step size $\Delta = 0.1°C$ were used and 33 iterations of steps were determined. Figure 5a shows the extracted regional maxima during the iterations which confirms the hypothesis that delamination is a subset of regional maxima. On the other hand, the gradient map was generated, cleaned, and clustered into two groups (K=2) to present background and delamination (Fig. 11a). Then two statistical measures were calculated from the clustered gradient map of delamination ($D_g$). Figure 5 (b and c) shows the intermediate results after screening by the two criteria during iteration.

**Table 1. Camera configurations**

| Case | Camera | Camera sensitivity | Image size | Spatial resolution (per pixel) | Survey height (m) |
|---|---|---|---|---|---|
| Slab | FLIR thermal | 0.02°C | 817x720 | ~0.14 cm | ~2.6 |
| Bridge | FLIR thermal | 0.02°C | 447x2987 | ~2.17 cm | ~53 |

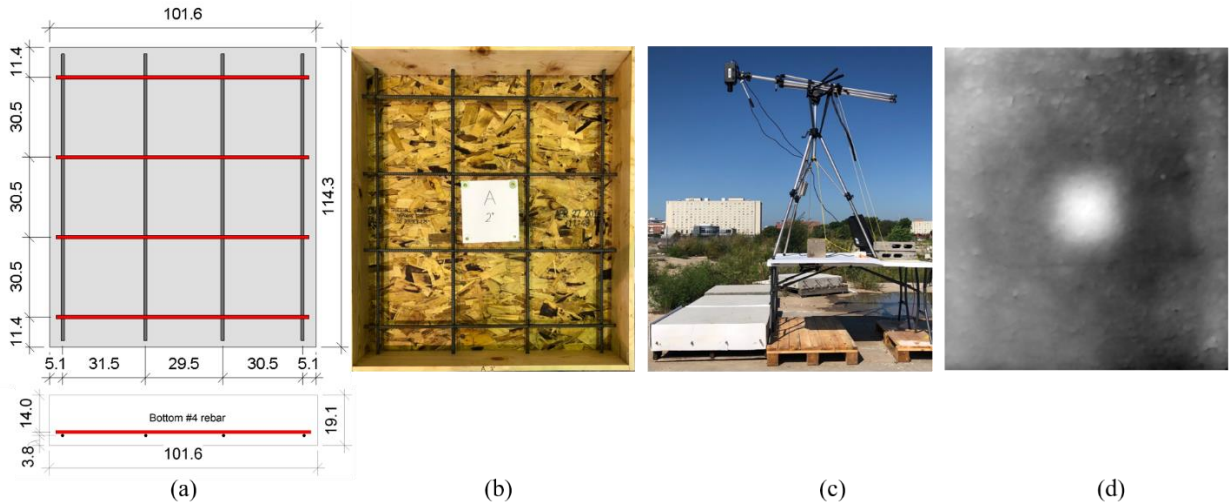

**FIGURE 4.** Experimental setup: (a)dimension and layout of reinforced concrete (cm); (b)the simulated delamination with the size of 25 by 25 cm and depth of 4.5 cm from the top surface; (c) data collection; (d) raw thermal image at 3 pm

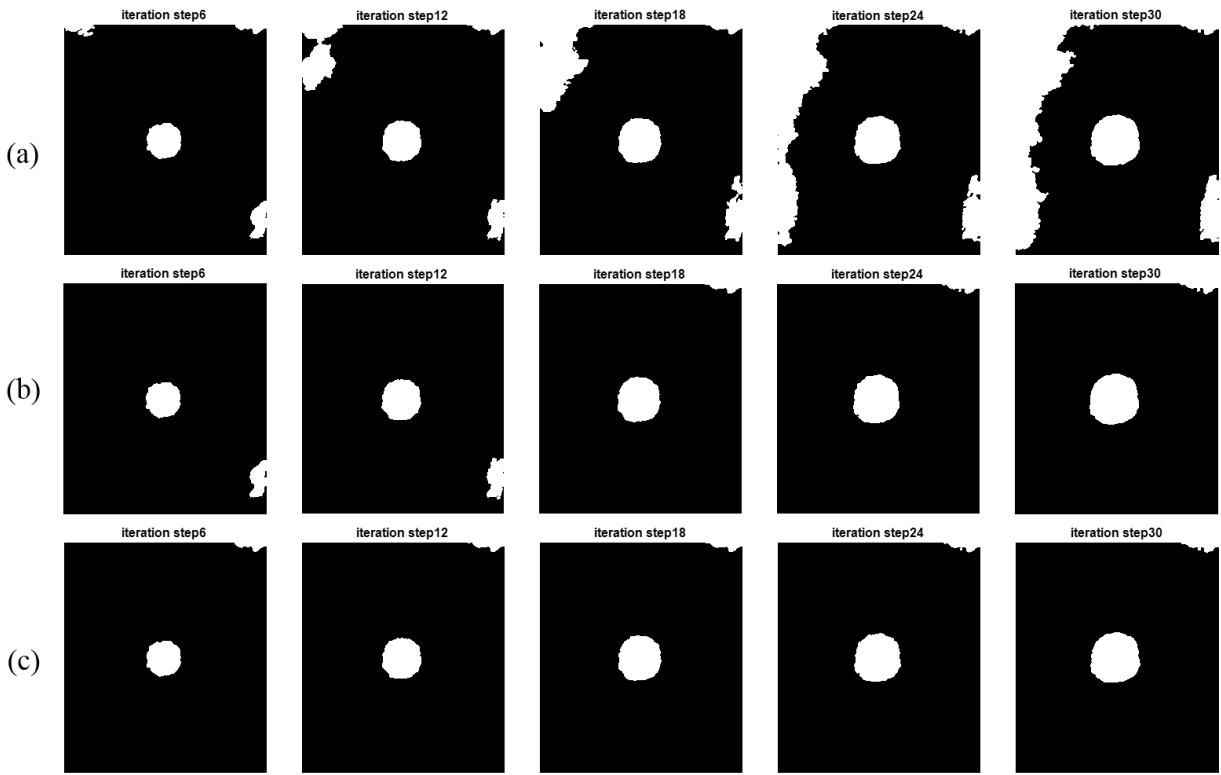

**FIGURE 5.** intermediate results during iteration: (a) detected regional maxima; (b) screened based on gradient mean $[M_{grad} - \frac{\delta_{std}}{2}, M_{grad} + \frac{\delta_{std}}{2}]$; (c) screened based on coefficient of variation $[0.5V_{var}, 1.9V_{var}]$.

The segmentation was then achieved through combining the screened regional maxima from each iteration under union operation. The outcome is shown in Fig. 6b to compare to existing methods (Oh et al. 2012; Omar and Nehdi 2017) whose outcomes are shown in Fig. 6(c d and e). The k-means clustering (k=2) segmented the image into the background and foreground (potential delamination) shown in Fig. 6c. The size and shape of simulated delamination were preserved while a large area of misclassification occurred. The threshold method which directly applied on the raw temperature suffered similar performance. When setting a lower threshold (Fig. 6d), the shape and size of delamination stayed the same but more misclassified areas were identified. When setting a higher threshold (Fig. 6e), the size was not preserved. The proposed method outperformed the threshold-based and clustering-based methods in terms of shape preservation and accuracy.

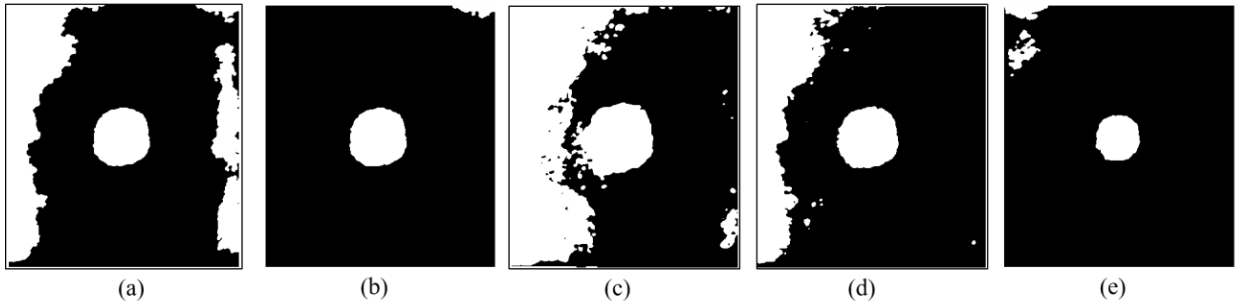

**FIGURE 6. Results and comparison: (a) total regional maxima detected, (b) proposed method after applying discrimination criteria (c) K-mean clustering with k=2, (d) threshold > 46.3°C, (e) threshold>47°C**

### 5. FIELD TEST
**5.1 Setup, Data Collection, and Result**

A field test was conducted on an in-service concrete bridge deck with a 10.16 cm thick concrete overlay at the northbound of US 77 close to Lincoln, Nebraska. The data was collected through UAVs equipped with optical and infrared cameras in October 16, 2017. Two configurations were implemented in Fig. 7 based on different payload requirements. DJI Matrice 600 was used to carry the infrared camera integrated with an onboard computer for thermal imaging. DJI Inspire One carried the optical camera for imaging the surface condition of the bridge. The path of UAV was designed to follow the center of the bridge at a fixed height and constant speed. The thermal image was then post-processed through the customized MATLAB algorithms for perspective correction, rotation, and stitching. Then the accuracy of the stitched thermal image was checked and calibrated through manual comparison with the visible image for this case.

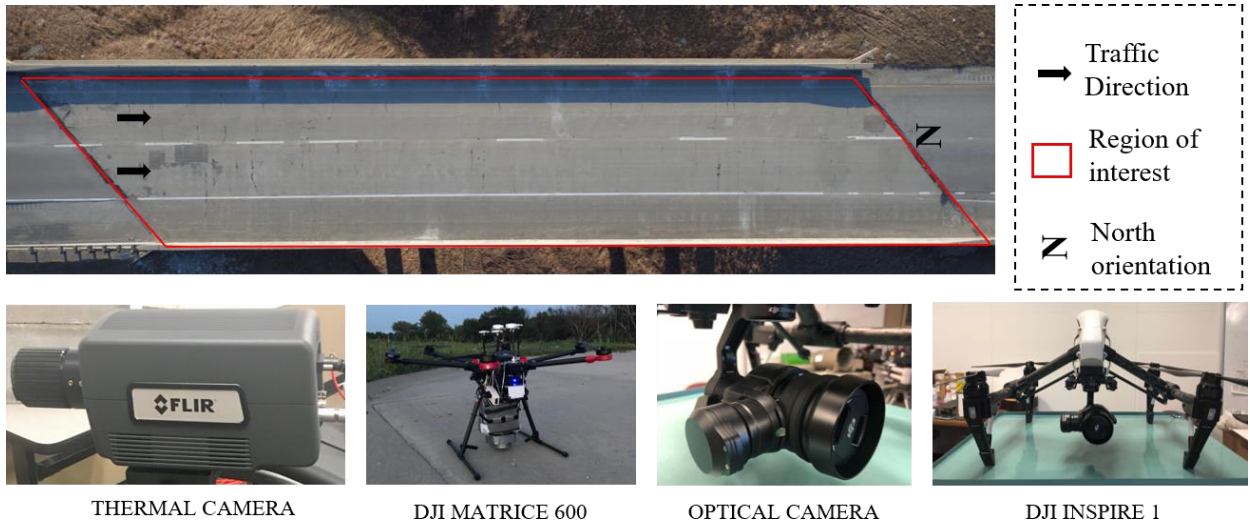

**FIGURE 7. Field setup and data collection**

Figure 8a is the raw thermal image of the bridge deck. The temperature of the bridge was color-coded where high to low temperature were represented through red to blue. The temperature variation was observed across the bridge. The highest temperature observed was about 33 °C which was around at the joint over the abutment while the lowest was around 19 °C for the top shadow. Since the data was collected in the afternoon, the potential delamination appeared as hot regions. The inspected regions had a temperature variation from 30.7 °C to 27.3 °C. Besides the hot regions, the potential sound area also had a temperature variation from 27.5 °C to 26 °C. This phenomenon had been widely observed in previous studies where the global threshold methods suffered. Figure 8b shows the total regional maxima identified throughout all iterations and Figure 8c shows the segmentation result by the proposed method where the initial offset $h_{in} = 0.5°C$ and step size $\Delta = 0.15°C$ were used and maximum 99 steps resulted in total iterations.

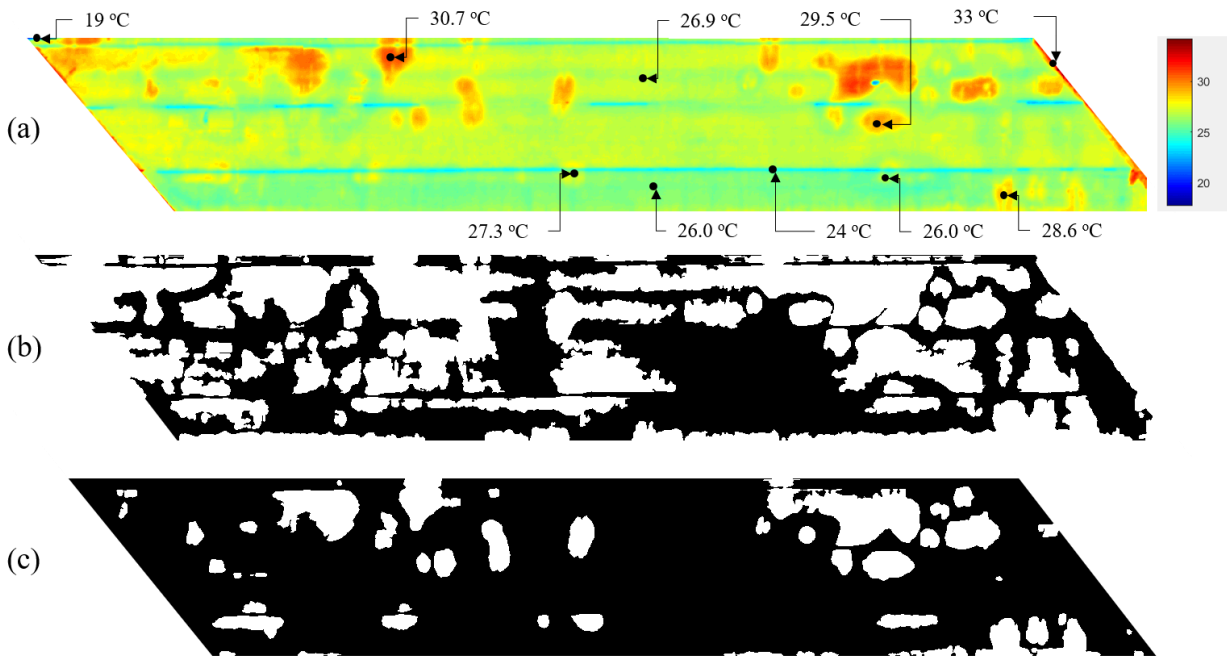

**FIGURE 8.** (a) thermal raw data; (b) total regional maxima detected; and (c) proposed method after applying discrimination criteria of gradient mean $[M_{grad} - \frac{\delta_{std}}{2}, M_{grad} + \frac{\delta_{std}}{2}]$ and coefficient of variation $[0.5V_{var}, 1.9V_{var}]$.

## 5.2 Validation with Hammer Sounding Test and Coring Samples, and Performance Comparison to Other Methods

Figure 9 (a) is the hammer sounding result provided by Nebraska Department of Transportation (NDOT) which is a trusted method for shallow delamination detection on site (ASTM 4850M-12) and the boundaries of delamination were directly marked with chalk lines while hammering. Both results from hammer sounding and the proposed method were further validated by coring samples (Fig. 9b). A total of 12 samples were cored on the bridge by NDOT where the location is marked in Fig. 9b. The results (Fig. 9b) of hammer sounding, the proposed method, and the coring samples agreed with each other: the debonding was found at the bottom of the concrete overlay which has a thickness around 6 cm (such as sample 1, 4, 7, 9 and 12). Especially in location 4, which had been missed by conventional methods (Fig. 9bcd), but was detected by the proposed method and confirmed the effect of non-uniform back variation.

The performance of global threshold and K-means clustering are compared against the hammer sounding result. Figure 9c shows the overestimation with the threshold of 80 percentile (27.3 ºC) of the maximum temperature where the bottom-half image had the sufficient indication of delamination, while the top-half image shows the overestimations made by this method. Figure 9d shows the global threshold method with the 82 percentile (28 ºC) and an underestimation of delamination occurred where the top-half image had a similar pattern to the hammer sounding result, but the bottom-half suffered the insufficient estimation. The K-means clustering result is showed in Fig. 9e which shows a closer estimation result in Fig. 9c. The proposed method (Figure 9b) compared to the other two methods, performed better in terms of the tolerance to the non-uniform background variation.

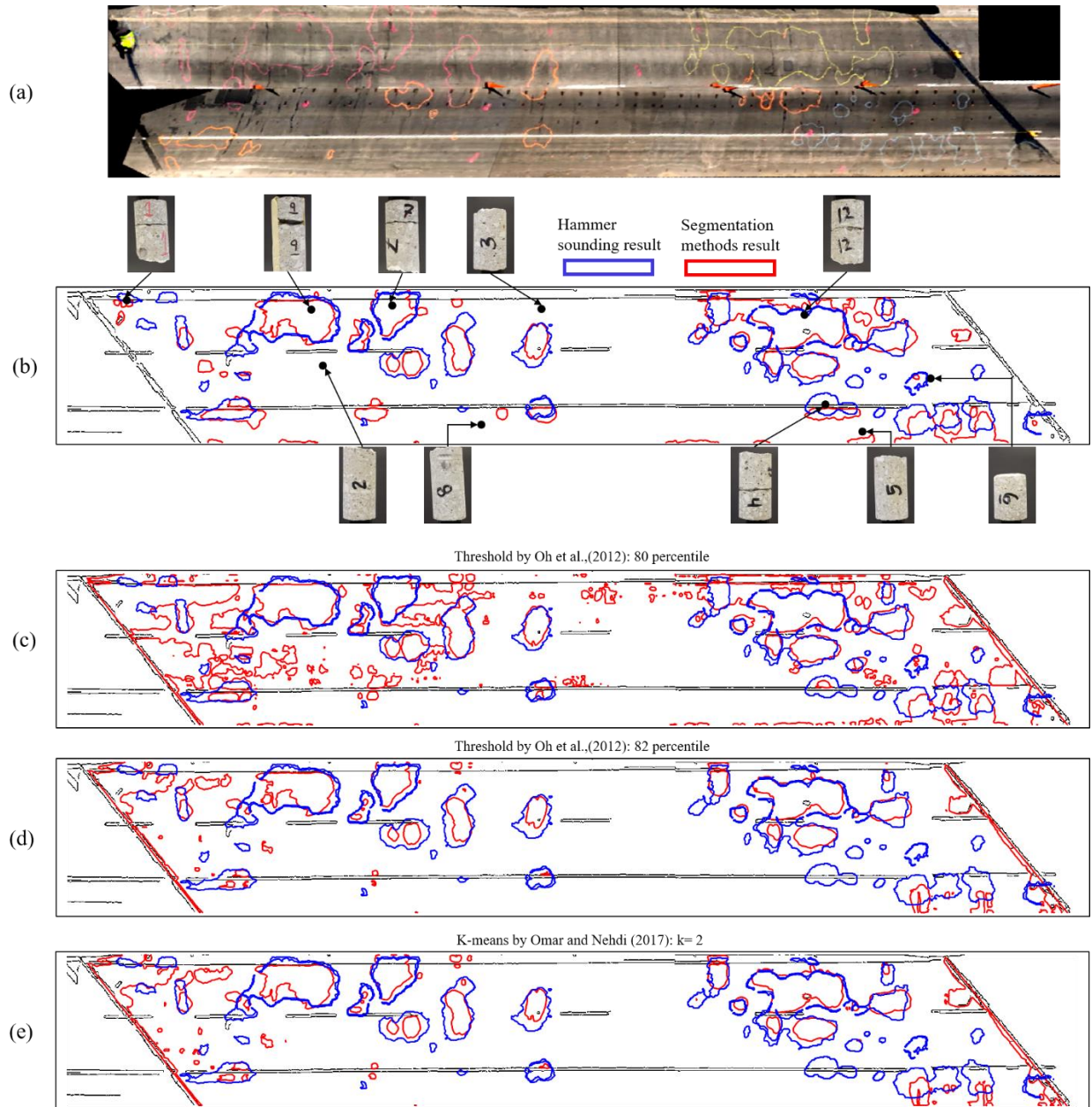

**FIGURE 9.** (a) hammer sounding results; (b) proposed method with coring validation; (c) 80 percentile threshold (27.3°C); (d) 82 percentile threshold (28°C); and (e) K-mean clustering with k=2.

## 6. DISCUSSIONS
### 6.1 Parameter Selection for Regional Maxima Extraction
There are two parameters for regional maxima extraction: the initial offset $h_{in}$ and step size $\Delta$. The initial offset aims to introduce the minimum contrast criterion recommended in the literature as the start point to distinguish delamination and sound area. The 0.5 °C value used in the study for both slab and bridge cases is within the recommended 0.4 °C to 0.6 °C range. The minimum step size of 0.05 is used based on maximum temperature sensitivity of the thermal camera. With the smallest step size, the finest expanding of regional maxima would be obtained so that more regional maxima could be preserved during iteration which gives the most sufficient extraction of regional maxima. With the increase of step size, the area of total regional maxima differs from the area of the finest step size (Fig. 10a), but also the maximum step required for iteration is decreased which give the efficiency of processing (Fig.10b). The step size less than 0.2 is recommended which would yield 10% difference in area compared to the finest step size.

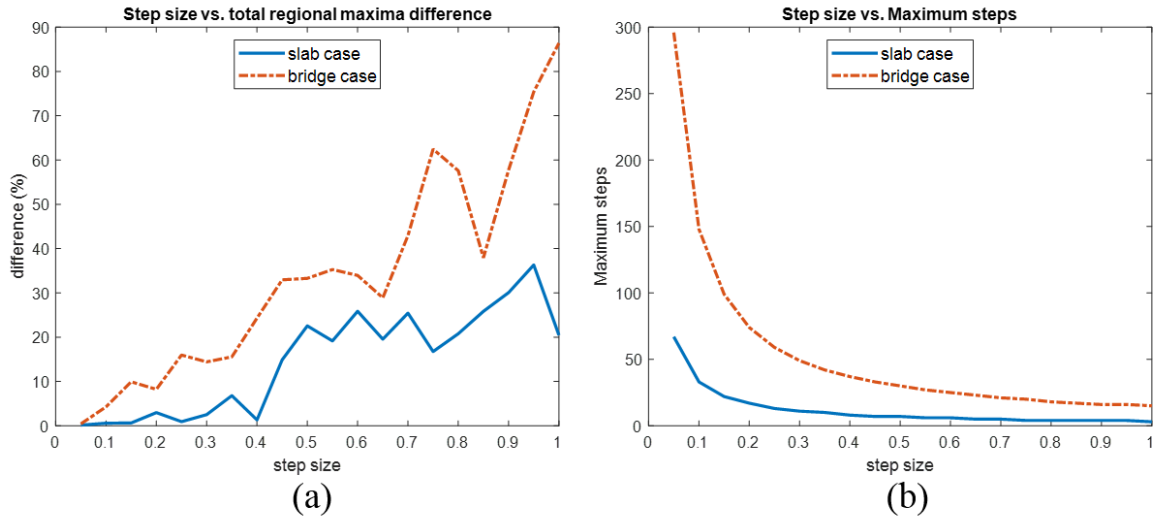

(a)  (b)

**FIGURE 10. Step size selection: (a) step size vs. total regional maxima difference by area; (b) step size vs. maximum step**

**6.2 Parameter Selection for Regional Maxima Discrimination**
Once the regional maxima have been determined, the measurement of the feature is needed to discriminate delamination out of regional maxima. We utilized the gradient information and its statistical measures as the criteria for distinguishing (Fig. 11). The generation of the gradient map followed a flux function proposed by Weickert (1998) with the sigma of 3.4 for slab case and 1.7 for bridge case. Size filtering was used to clean up insignificant small spots related to surface texture. Edge detection was used to remove areas of road paint. K-means clustering algorithm with k=2 was used to estimate the mask of boundaries of delamination (Fig. 11 a and b). Thus, the accuracy of delamination boundaries will affect the measurement of distributions. Here the thresholds of two metrics (gradient mean $[M_{grad} - \frac{\delta_{std}}{2}, M_{grad} + \frac{\delta_{std}}{2}]$ and coefficient of variation $[0.5V_{var}, 1.9V_{var}]$) was tuned based on experimental studies and a single real world case. The reliability needs to be evaluated by further studies on scenarios of different depth and weather conditions. Also, the rationale of using the threshold of gradient magnitude was based on the assumption that delamination in the surveyed image was at a relative same level of depth (case in this paper). However, delamination may occur at different level of depth which may cause the gradient different between groups and thus limits the applicability of the discrimination criterion proposed.

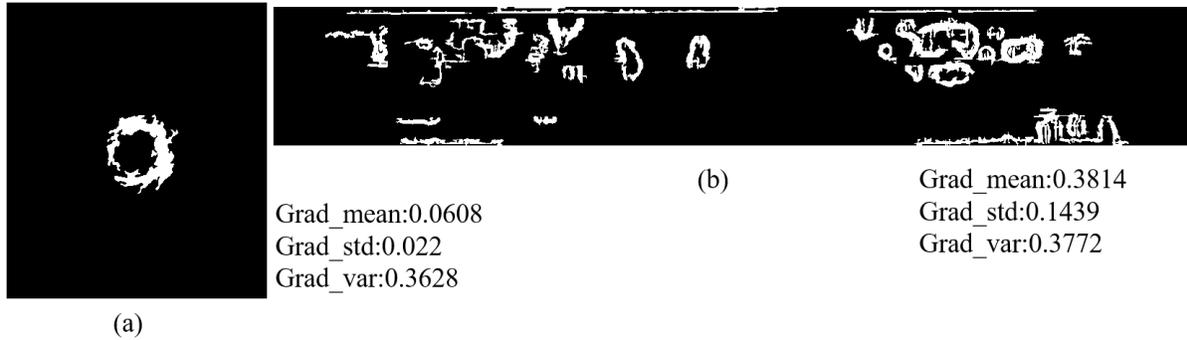

FIGURE 11. Gradient information of delamination: (a) mask of delamination in the slab with the mean of 0.0608, standard deviation of 0.022, and total variance of 0.3628; (b) mask of delamination in the bridge with the mean of 0.3814, standard deviation of 0.1439, and total variance of 0.3772

## 7. CONCLUSIONS

This paper proposed a top-down approach to segment delamination in concrete bridge decks using aerial thermal images to overcome issues of non-uniform temperature distribution at the background. The proposed method eliminates the need of inferencing the background required by existing methods. A weight decay function was used to well regularize the grayscale morphological reconstruction for regional maxima extraction. It has been observed that delamination areas in a thermal image is a subset of general regional maxima in mathematic morphology. Both experimental and field tests confirmed the observation and revealed that delamination could be discriminated through metrics from gradient information of the boundary. The improved performance was observed in both cases when compared to the conventional methods.

**ACKNOWLEDGMENT**
The authors would like to thank Nebraska Department of Transportation for their efforts in facilitating the data collection and sharing their non-destructive evaluation results.

# REFERENCES

Abdel-Qader, I., Yohali, S., Abudayyeh, O., and Yehia, S. (2008). "Segmentation of thermal images for non-destructive evaluation of bridge decks." *Ndt & E International*, 41(5), 395-405.

Cheng, C., Na, R., and Shen, Z. (2018). "Thermographic Laplacian-Pyramid Filtering to Enhance Delamination Detection in Concrete Structure." *Infrared Physics & Technology*.

Cheng, C., and Shen, Z. "Time-Series Based Thermography on Concrete Block Void Detection." *Proc., Construction Research Congress 2018*, 732-742.

Cheng, L., Zhao, W., Han, P., Zhang, W., Shan, J., Liu, Y., and Li, M. (2013). "Building region derivation from LiDAR data using a reversed iterative mathematic morphological algorithm." *Optics Communications*, 286, 244-250.

Coster, M., and Chermant, J.-L. (2001). "Image analysis and mathematical morphology for civil engineering materials." *Cement and Concrete Composites*, 23(2-3), 133-151.

Dabous, S. A., Yaghi, S., Alkass, S., and Moselhi, O. (2017). "Concrete bridge deck condition assessment using IR Thermography and Ground Penetrating Radar technologies." *Automation in Construction*.

Ellenberg, A., Kontsos, A., Moon, F., and Bartoli, I. (2016). "Bridge Deck delamination identification from unmanned aerial vehicle infrared imagery." *Automation in Construction*, 72, 155-165.

Haralick, R. M., Sternberg, S. R., and Zhuang, X. (1987). "Image analysis using mathematical morphology." *IEEE transactions on pattern analysis and machine intelligence*(4), 532-550.

Hiasa, S., Birgul, R., and Catbas, F. N. (2017). "Effect of defect size on subsurface defect detectability and defect depth estimation for concrete structures by infrared thermography." *Journal of Nondestructive Evaluation*, 36(3), 57.

Hiasa, S., Birgul, R., and Catbas, F. N. (2017). "Investigation of effective utilization of infrared thermography (IRT) through advanced finite element modeling." *Construction and Building Materials*, 150, 295-309.

Kee, S.-H., Oh, T., Popovics, J. S., Arndt, R. W., and Zhu, J. (2011). "Nondestructive bridge deck testing with air-coupled impact-echo and infrared thermography." *Journal of Bridge Engineering*, 17(6), 928-939.

Maragos, P. (1987). "Tutorial on advances in morphological image processing and analysis." *Optical engineering*, 26(7), 267623.

Matas, J., Chum, O., Urban, M., and Pajdla, T. (2004). "Robust wide-baseline stereo from maximally stable extremal regions." *Image and vision computing*, 22(10), 761-767.

Milovanović, B., Banjad Pečur, I., and Štirmer, N. (2017). "The methodology for defect quantification in concrete using IR thermography." *Journal of Civil Engineering and Management*, 23(5), 573-582.

Oh, T., Kee, S.-H., Arndt, R. W., Popovics, J. S., and Zhu, J. (2012). "Comparison of NDT methods for assessment of a concrete bridge deck." *Journal of Engineering Mechanics*, 139(3), 305-314.

Omar, T., and Nehdi, M. L. "Clustering-Based Threshold Model for Condition Assessment of Concrete Bridge Decks Using Infrared Thermography." *Proc., International Congress and Exhibition" Sustainable Civil Infrastructures: Innovative Infrastructure Geotechnology"*, Springer, 242-253.

Omar, T., and Nehdi, M. L. (2017). "Remote sensing of concrete bridge decks using unmanned aerial vehicle infrared thermography." *Automation in Construction*, 83, 360-371.

Rishikeshan, C., and Ramesh, H. (2018). "An automated mathematical morphology driven algorithm for water body extraction from remotely sensed images." *ISPRS journal of photogrammetry and remote sensing*, 146, 11-21.

Sultan, A. A., and Washer, G. (2017). "A pixel-by-pixel reliability analysis of infrared thermography (IRT) for the detection of subsurface delamination." *NDT & E International*, 92, 177-186.

Sun, B.-C., and Qiu, Y.-j. "Automatic identification of pavement cracks using mathematic morphology." *Proc., International Conference on Transportation Engineering 2007*, 1783-1788.

Vaghefi, K., Ahlborn, T. M., Harris, D. K., and Brooks, C. N. (2013). "Combined imaging technologies for concrete bridge deck condition assessment." *Journal of Performance of Constructed Facilities*, 29(4), 04014102.

Vincent, L. (1993). "Morphological grayscale reconstruction in image analysis: applications and efficient algorithms." *IEEE transactions on image processing*, 2(2), 176-201.

Vosselman, G. (2000). "Slope based filtering of laser altimetry data." *International Archives of Photogrammetry and Remote Sensing*, 33(B3/2; PART 3), 935-942.

Washer, G., Fenwick, R., Bolleni, N., and Harper, J. (2009). "Effects of environmental variables on infrared imaging of subsurface features of concrete bridges." *Transportation Research Record: Journal of the Transportation Research Board*(2108), 107-114.

Watase, A., Birgul, R., Hiasa, S., Matsumoto, M., Mitani, K., and Catbas, F. N. (2015). "Practical identification of favorable time windows for infrared thermography for concrete bridge evaluation." *Construction and Building Materials*, 101, 1016-1030.

Wei, N., Zhao, X., Wang, T., and Song, H. "Mathematical morphology based asphalt pavement crack detection." *Proc., International Conference on Transportation Engineering 2009*, 3883-3887.

Weickert, J. (1998). *Anisotropic diffusion in image processing*, Teubner Stuttgart.